\documentclass[twocolumn,aps,prl,superscriptaddress,showpacs]{revtex4-1}
\usepackage[usenames,dvipsnames]{color}

\usepackage{amsmath,amsfonts,amsthm}

\usepackage{graphicx,color}
\usepackage{multirow}
\usepackage{bbm}
\usepackage{textcomp}
\usepackage[utf8]{inputenc}
\DeclareMathOperator{\Proj}{proj}
\DeclareMathOperator{\Span}{span}

\usepackage{amsmath,amsfonts,amssymb,amscd}
\usepackage{graphicx}

\newcommand{\bra}[1]{\ensuremath{\langle#1|}}

\newcommand{\ket}[1]{\ensuremath{|#1\rangle}}

\newcommand{\braket}[1]{\ensuremath{\langle #1 \rangle}}

\newcommand{\id}{\ensuremath{\mathbbm 1}}

\newcommand{\BB}{\ensuremath{\mathcal{B}}}

\newcommand{\BE}{\begin{equation}}
\newcommand{\EE}{\end{equation}}
\newcommand{\be}{\begin{equation}}
\newcommand{\ee}{\end{equation}}
\newcommand{\bea}{\begin{eqnarray}}
\newcommand{\eea}{\end{eqnarray}}
\newcommand{\bean}{\begin{eqnarray*}}
\newcommand{\eean}{\end{eqnarray*}}

\newcommand{\mean}[1]{\ensuremath{\langle #1 \rangle}}

\newcommand{\tr}{{\rm Tr}}

\newcommand{\bc}{\begin{center}}
\newcommand{\ec}{\end{center}}

\renewcommand{\vr}{\ensuremath{\varrho}} 

\newcommand{\kommentar}[1]{}

\begin{document}

\title{Bounding Temporal Quantum Correlations}

\author{Costantino Budroni} 
\author{Tobias Moroder} 
\author{Matthias Kleinmann}
\author{Otfried G{\"u}hne} 
\affiliation{Naturwissenschaftlich-Technische Fakult\"at,
Universit\"at Siegen,
Walter-Flex-Straße~3,
D-57068 Siegen, Germany}

\date{\today}

\begin{abstract}
Sequential measurements on a single particle play an important role in 
fundamental tests of quantum mechanics. We provide a general method to analyze temporal quantum 
correlations, which allows us to compute the maximal correlations for sequential 
measurements in quantum mechanics. As an application, we present the full characterization of temporal correlations in the simplest Leggett-Garg scenario and in the sequential measurement scenario associated with the most fundamental 
proof of the Kochen-Specker theorem.
\end{abstract}


\maketitle

{\it Introduction.---}The physics of microscopic systems is governed by the laws of quantum
mechanics and exhibits many features that are absent in the classical 
world. The best-known result showing such a difference is due to
Bell \cite{Bell64}. The assumptions of realism and locality 
lead to bounds on the correlations---the Bell inequalities, and 
these bounds are violated in quantum mechanics. Interestingly, this 
quantum violation is limited for many Bell inequalities and does not 
reach the maximal possible value. For instance, the Bell inequality 
derived by Clauser, Horne, Shimony, and Holt (CHSH) bounds 
the correlation \cite{CHSH69}
\begin{equation}\label{chsh}
\BB=\mean{A_1 \otimes B_1}+\mean{A_1 \otimes B_2}
+\mean{A_2 \otimes B_1}-\mean{A_2 \otimes B_2},
\end{equation}
 where  $A_i$ and  $B_j$ are measurements on two different particles.
On the one hand, local realistic models obey the CHSH inequality
$\BB \leq 2$, which is violated in quantum mechanics.
On the other hand, the maximal quantum value is upper bounded by $
\BB\le 2\sqrt{2}$, a result known as Tsirelson's bound \cite{Tsirelson80}.
Whereas this bound holds within quantum mechanics, it has turned out 
that hypothetical theories that reach the algebraic maximum 
$\BB=4$ without allowing faster-than-light communication
are possible \cite{prbox}. This raises the question of whether the 
bounded quantum value can be derived on physical grounds from fundamental 
principles. Partial results are available, and principles have been suggested 
that bound the correlations: in a world where maximal correlations are 
observed, the communication complexity is trivial \cite{BBLMTU06}, a 
principle established as information causality is violated \cite{PPKSWZ09}, 
and there exists no reversible dynamics \cite{gross}.

The question of how and why quantum correlations are fundamentally
limited has been discussed mainly in the scenario of bipartite 
and multipartite measurements. What happens, however, if we shift 
the attention from spatially separated measurements to temporally 
ordered measurements? There is no need to measure on distinct systems 
as in Eq.~\eqref{chsh}, but rather, we may perform sequential measurements 
on the same system. Then, an elementary property of quantum 
mechanics becomes important: the measurement changes the state of 
the system. In fact, this allows us to temporally ``transmit'' a 
certain amount of information \cite{Fritz10}, and one would expect 
that the correlations in the temporal case can be larger than in 
the spatial situation.

\begin{figure*}[t]
\includegraphics[width=0.65\textwidth,viewport=550 900 1800 
1500]{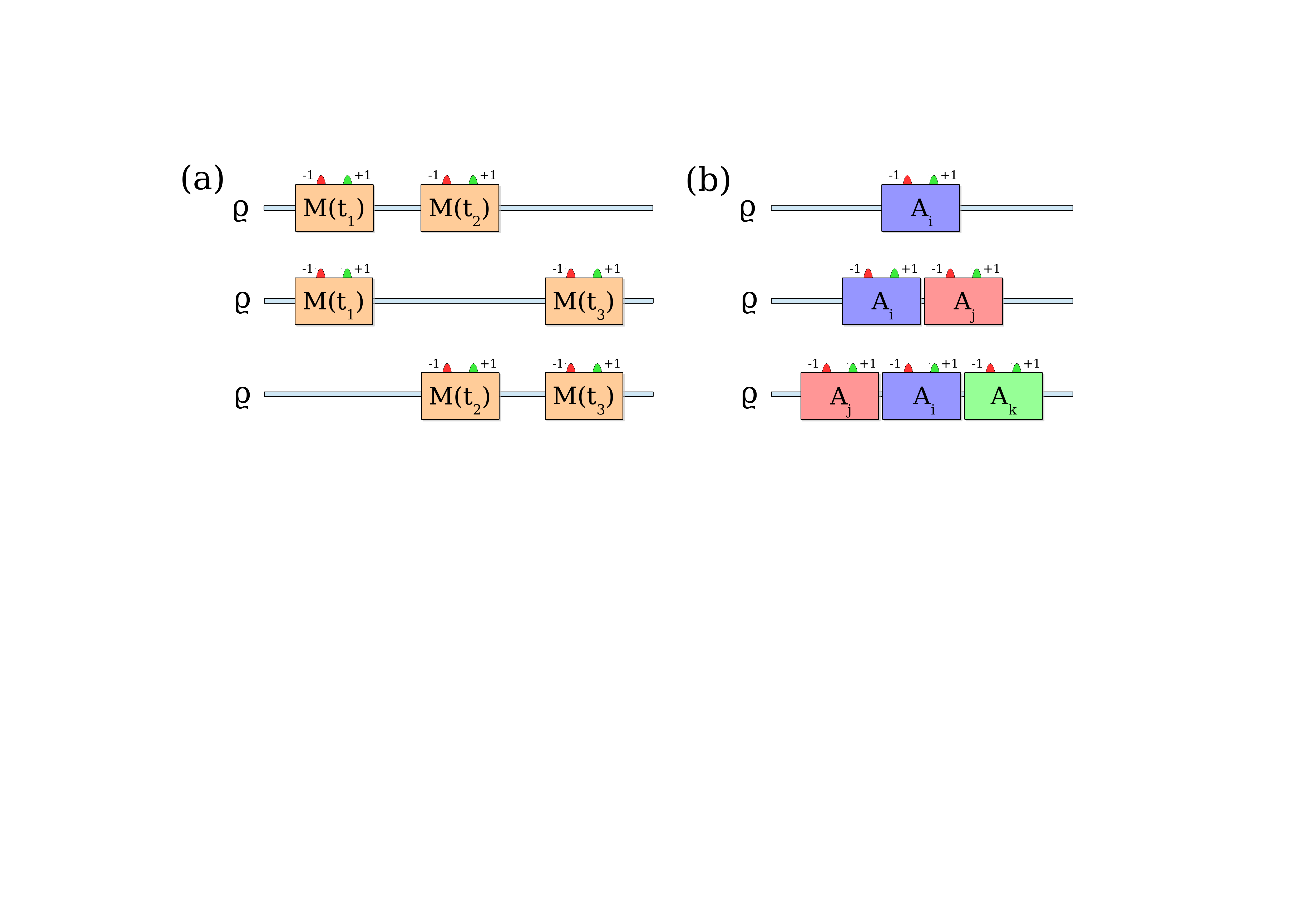}
\caption{\label{fig1:situation} Sequential measurements
occur in two different scenarios. (a) In the Leggett-Garg scenario,
one takes a single observable $M$ that measures whether
the physical system is in one of two possible macroscopic
states. Then, one considers the correlations between these
measurements at three different times,
$\mean{M(t_1) M(t_2)}_{\rm seq}$,
$\mean{M(t_1) M(t_3)}_{\rm seq}$,
and
$\mean{M(t_2) M(t_3)}_{\rm seq}$.
The values predicted by quantum mechanics contradict the assumption
that the physical system is in any of these macroscopic states at
any time and that the measurement reveals this state without disturbing
it.
(b) In the Kochen-Specker scenario one considers a set of observables
$A_i$. Some of these observables are compatible and can therefore be measured 
simultaneously or in a sequence without any disturbance. Then one measures 
the correlations of simultaneous or sequential measurements of compatible 
observables, such as
$\mean{A_i}_{\rm seq}$,
$\mean{A_i A_j}_{\rm seq}$,
and
$\mean{A_j A_i A_k}_{\rm seq}$.
For these correlations, one finds that quantum mechanics contradicts the
assumption of noncontextuality. This assumption states that the result of
a measurement should not depend on which other compatible observables are
measured along with it. It should be noted, however, that the 
situation considered in this Letter is more general than  case (a) or (b), 
since no assumption about the time evolution or the compatibility of 
observables is made.
}
\end{figure*}

We stress that sequential measurements also have been considered in
the analysis of the question how quantum and classical mechanics are
different, the most well-established results here are quantum contextuality 
(the Kochen-Specker theorem \cite{KochenSpecker67}) and macrorealism 
(Leggett-Garg inequalities \cite{LeggettGarg85}); cf.\ Fig.~1.
The research in this fields has triggered experiments involving
sequential measurements. For demonstrating such a contradiction between 
classical and quantum physics, e.g., the correlation 
\begin{align}
\mathcal{S}_5  = &\mean{A_1A_2}_{\rm seq}+\mean{A_2A_3}_{\rm
seq}+\mean{A_3A_4}_{\rm seq}
+\mean{A_4A_5}_{\rm seq} 
\nonumber
\\
&- \mean{A_5A_1}_{\rm seq}
\label{klyachko}
\end{align}
 has been considered \cite{KCBS08, Barbieri09}. Here, $\mean{A_iA_j}_\mathrm{seq}$ 
 denotes  a sequential expectation value that is the average of the product of the 
 value of the observables $A_i$ and $A_j$ when first $A_i$ is measured, and afterwards 
 $A_j$. One can show that for macrorealistic theories as well as for
 noncontextual models the bound $\mathcal{S}_5\leq 3$ holds, but in quantum 
 mechanics, this can be violated.

Here however, we are rather interested in the fundamental bounds on the
temporal quantum correlations, with no assumption about the compatibility of the observables. Special cases of this problem have been 
discussed before: for Leggett-Garg inequalities, maximal values for two-level 
systems have been derived \cite{Barbieri09, KoflerBrukner08}, and temporal
inequalities similar to the CHSH inequality have been discussed 
\cite{Fritz10,Fritz09}.

We provide a method that allows us to compute the maximal achievable
quantum value for an arbitrary inequality and thus we solve the 
problem of bounding temporal quantum correlations. First, we will 
discuss a simple method, which can be used for expressions as in 
Eq.~(\ref{klyachko}), where only sequences of two measurements are
considered. Then, we introduce a general method which can be used 
for arbitrary sequential measurements, resulting in a complete 
characterization of the possible quantum values. Interestingly, 
our methods characterize temporal correlations exactly, whereas for 
the case of spatially separated measurements only converging
approximations are known.

{\it Projective measurements.---}When determining the maximal value for sequential measurements as 
in Eq.~(\ref{klyachko}) we consider projective measurements, as 
these are the standard textbook examples of quantum measurements. 
The underlying formalism has been established by von Neumann \cite{neumann} 
and L\"uders \cite{lueders}. An observable $A$ with possible results 
$\pm 1$ is described by two projectors $\Pi_+$ and $\Pi_-$ such that
$A = \Pi_+ - \Pi_-$. If the observable $A$ is measured, the quantum 
state is projected onto the space of the observed result, i.e., 
$\varrho \mapsto \Pi_\pm \varrho\Pi_\pm/\tr(\varrho \Pi_\pm).$ 
Applying this scheme to the case of sequential measurements, one finds 
that the sequential mean value can be written as 
\begin{equation}
\mean{A_iA_j}_{\rm seq}  = \frac{1}{2} [\tr(\varrho A_i A_j ) + \tr(\varrho A_j A_i )].
\label{sequence}
\end{equation}
It is interesting to notice that for pairs of $\pm 1$-valued observables 
such a mean value does not depend on the order of the measurement \cite{Fritz10}.

{\it The simplified method.---}We first show how the maximal quantum mechanical value for an expression 
such as $\mathcal{S}_5$ in Eq.~(\ref{klyachko}) can be determined. First, we 
consider a  set $\mathcal{A}=\{A_i\}$ of $\pm 1$-valued observables and 
a general expression $C=\sum_{ij} \lambda_{ij}\mean{A_i A_j}_{\rm seq}$. 
The correlations given in Eq.~(\ref{klyachko}) are just a special case 
of this scenario. Then, we consider the matrix built up by the sequential 
mean values $X_{ij}=\mean{A_i A_j}_{\rm seq}$. This matrix has the 
following properties: (i) it is real and symmetric, $X=X^T$, (ii) the 
diagonal elements equal one, $X_{ii}=1$, and (iii) the matrix has 
no negative eigenvalue (or $v^T X v \geq 0$ for any 
vector $v$), denoted as $X \succeq 0$ (see Appendix~A2). 
A similar construction for the matrix $X$, together with the optimization 
problem below, has been considered before in relation with Bell 
inequalities \cite{WEH06}. However, our method involves a different 
notion of correlations, namely that given by Eq.~(\ref{sequence}).

The main idea is now to optimize the expression ${C=\sum_{ij} \lambda_{ij} X_{ij}}$ 
over all matrices with the properties (i)--(iii) above. Hence, we consider 
the optimization problem
\begin{eqnarray}\label{eq:sdp}
\mbox{maximize:} && \sum_{ij} \lambda_{ij} X_{ij},
\\ \nonumber 
\mbox{subjected to:} && X=X^T \succeq 0 \mbox{ and for all $i$, } 
X_{ii} = 1.
\label{genopt}
\end{eqnarray}
Since all matrices $X$ that can originate from a sequence of quantum 
measurements will be of this form, one performs the optimization over 
a potentially larger set. Thus, the solution of this optimization is, 
in principle, just an upper bound on the maximal quantum value 
of $\mathcal{S}_5$.
Note that the optimization itself can be done efficiently and is 
assured to reach the global optimum since it represents a so-called 
semidefinite program~\cite{sdpref}. In the case of $\mathcal{S}_5$, 
this optimization can even be solved analytically and gives
\begin{equation}
 \mathcal{S}_5 \leq  \frac{5}{4}\left(1+\sqrt{5}\right) \approx 4.04.
\end{equation}
It turns out that appropriately chosen measurements on a qubit already 
reach this value (see Appendix A2 and Refs.~\cite{Barbieri09,dimwit}). Hence, 
this upper bound is tight. More generally, one can prove that each matrix 
$X$ with the above properties has a sequential quantum representation (see Appendix~A2).
Finally, note that if the observables in each sequence are required to 
commute, then the maximal quantum value for $\mathcal{S}_5$ is known to 
be $\Omega_{QM}= 4\sqrt{5}-5\approx 3.94$ \cite{LSW11,AQBTC12}.

{\it The general method.---}The above method can only be used for correlations terms of sequences of at most two $\pm 1$-valued observables. In the following, we discuss the conditions allowing a given probability distribution to be realized as sequences of measurements on a single quantum system in the general setting. We label as ${\bf r}=(r_1,r_2,\dots,r_n)$ the results of an $n$-length sequence obtained by using the setting ${\bf s}=(s_1,s_2,\dots,s_n)$. The ordering is such that $r_1,s_1$ label the result and the 
setting for the first measurement etc. The outcomes of any such sequence are sampled from the sequential conditional probability distribution 
\begin{equation}
P({\bf r | s}) \equiv P_{\rm seq}(r_1,r_2,\dots, r_n|s_1,s_2,\dots,s_n).
\end{equation}
In the case of projective quantum measurements, each individual result $r$ of any setting $s$ is associated with a projector $\Pi_r^s$, which altogether satisfy two requirements: for each setting the operators must sum up to the identity, i.e., $\sum_{r} \Pi_r^s = \mathbbm{1}$ and they satisfy the orthogonality relations  $\Pi_{r}^s \Pi_{r'}^s =\delta_{rr'} \Pi_{r}^s$, where $\delta_{rr'}$ is the Kronecker symbol. Finally, after the measurement with the setting $s$ and result $r$, the quantum state is transformed according to the rule $\varrho \mapsto \Pi_r^s \varrho \Pi_r^s/P(r|s)$. 

In the following, we say that the a conditional probability distribution $P({\bf r | s})$ has a sequential projective quantum representation if there exists a suitable set of such operators $\Pi_r^s$ and an appropriate initial state $\varrho$ such that  
\begin{equation}
P({\bf r|s}) = \tr[ \Pi({\bf r|s}) \Pi({\bf r|s})^\dag \varrho],
\end{equation}
with the shorthand $\Pi({\bf r|s}) = \Pi_{r_1}^{s_1} \Pi_{r_2}^{s_2} \cdots
\Pi_{r_n}^{s_n}$. 

Whether a given distribution $P({\bf r|s})$ indeed has such a representation can be answered via a so-called matrix of moments, which often appears in moment problems~\cite{WEH06,NPA07, NPA08,doherty08}. This matrix, denoted as $M$ in the following, contains the expectation value of the products of the above-used operators $\Pi({\bf r|s})$ at the respective position in the matrix. In order to identify this position we use as a label the abstract operator sequence ${\bf r|s}$ for both row and column index. In this way the matrix is defined as 
\begin{equation}
M_{{\bf r|s}; {\bf r^\prime|s^\prime}} = \braket{\Pi({\bf r|s}) \Pi({\bf r^\prime|s^\prime})^\dag}.
\end{equation}
Whenever this matrix is indeed given by a sequential projective quantum representation, the matrix $M$ satisfies two conditions: (a) linear relations of the form $M_{{\bf r|s; k|l}} = M_{{\bf r^\prime|s^\prime; k^\prime|l^\prime}}$ if the underlying operators are equal as a consequence of the properties of normalization and orthogonality of projectors, (b)  $M \succeq 0$ since $v^\dagger M v \geq 0$ holds
for any vector $v$, because such a product can be written as the expectation value $\braket{CC^\dag}_{\varrho}~\geq~0$  which is non-negative for any operator $C$. Finally, note that certain entries of this matrix are the given probability distribution, for instance, at the diagonal $M_{{\bf r|s; r|s}} = P({\bf r|s})$. The main point, however, is the converse statement:
given a moment matrix with properties (a) and (b) above, the associated probability distribution $P({\bf r|s})$ always has a sequential projective quantum representation (see Appendix~A3). 

Hence, the search for quantum bounds represents again a semidefinite program. The fact that this characterization is sufficient is in stark contrast with the analogue technique in the spatial Bell-type scenario~\cite{NPA07, NPA08}, where one needs to use moment matrices of an increasing size $n$ to generate better superset characterizations which only become sufficient in the limit $n \rightarrow \infty$. However, indirectly, the sufficiency of our method has already been proven in this context~\cite{NPA08} (see Appendix~A3).

\begin{figure}[t]
\includegraphics[width=0.9\columnwidth]{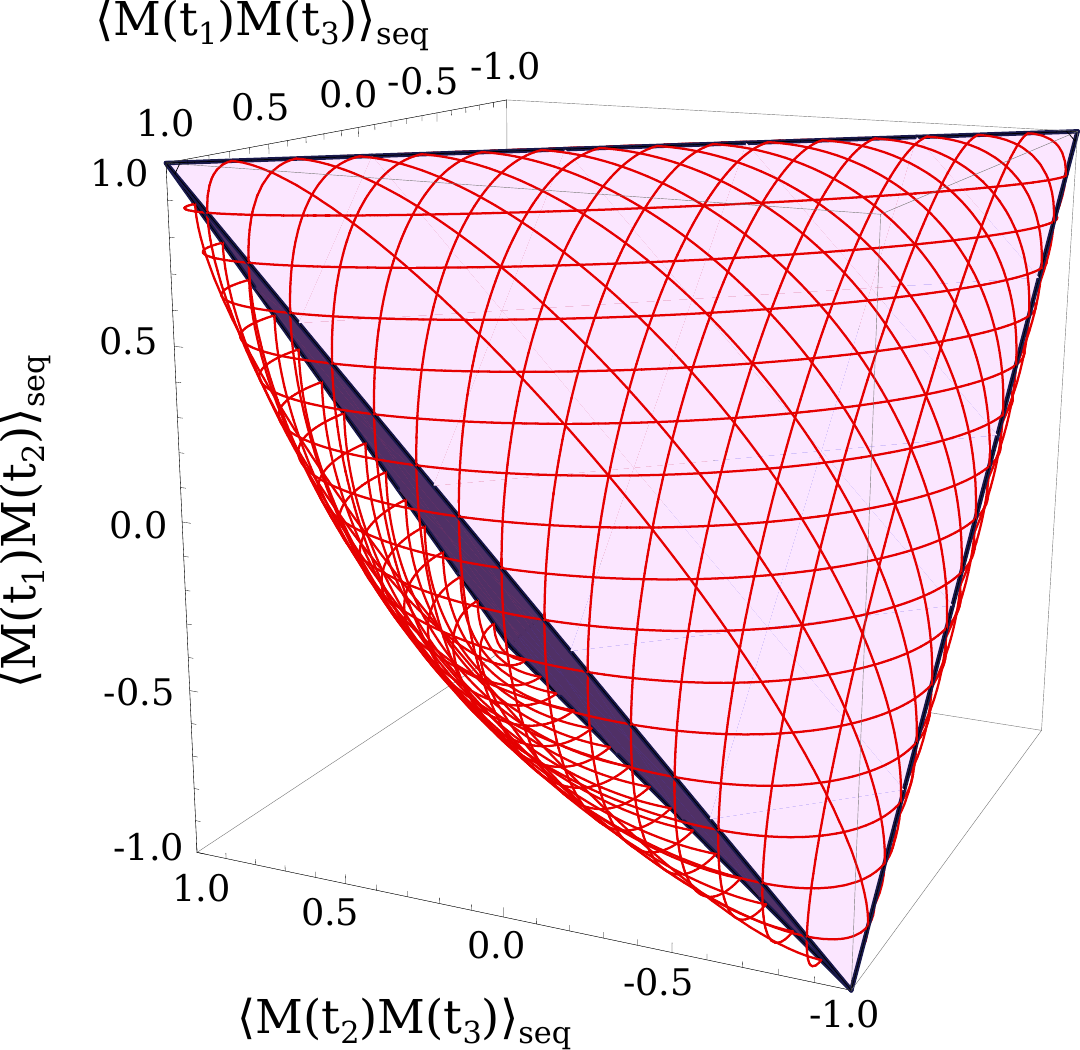}
\caption{\label{fig2:leggett} Complete characterization of the possible
quantum values  for the simplest Leggett-Garg scenario. In this case, 
three different times are considered, resulting in three possible correlations
$\mean{M(t_1) M(t_2)}_{\rm seq}$,
$\mean{M(t_1) M(t_3)}_{\rm seq}$,
and
$\mean{M(t_2) M(t_3)}_{\rm seq}$.
In this three-dimensional space, the possible classical values form a tetrahedron,
characterized by Eq.~(\ref{leggettsimple}) and variants thereof. The possible
quantum mechanical values form a strictly larger set with curved boundaries.
}
\end{figure}

{\it Applications.---}To demonstrate the effectiveness of our approach, 
we discuss four examples. First, we consider the 
original Leggett-Garg inequality
\begin{equation}
\begin{split}
\label{leggettsimple}
\mathcal{S} = 
\mean{M(t_1)M(t_2)}_{\rm seq}+\mean{M(t_2)M(t_3)}_{\rm seq} \\-
\mean{M(t_1)M(t_3)}_{\rm seq} \leq 1.
\end{split}
\end{equation}
This bound holds for macrorealistic models, and it has been shown 
that in quantum mechanics values up to $\mathcal{S}=3/2$ can be 
observed \cite{LeggettGarg85, Barbieri09, KoflerBrukner08}. Our 
methods allow us not only to prove that this value is optimal for 
any dimension and any measurement, but also to, for instance, determine all 
values in the three-dimensional space of temporal correlations 
$\mean{M(t_i) M(t_j)}$, which can originate from quantum mechanics. 
The detailed description is given in Fig.~2, and the
calculations are given in the Appendix~A1.

Second, we consider generalizations of the Eq.~(\ref{klyachko}) with 
a larger number of measurements, known as $N$-cycle inequality 
\cite{LSW11,AQBTC12},
\begin{equation}
\mathcal{S}_N = \sum_{i=0}^{N-2} \mean{A_i A_{i+1}}_{\rm seq}
-\mean{A_{N-1} A_0}_{\rm seq}.
\label{klyachkon}
\end{equation}
For this case, everything can be solved analytically (see Appendix A2)
leading to the bound
\begin{equation}
\mathcal{S}_N \leq N \cos\left(\frac{\pi}{N}\right),
\end{equation}
which can be reached by suitably chosen measurements. 
This value has already occurred in the literature \cite{Barbieri09,dimwit}, 
but only qubits have been considered. Our proof shows that it is valid in 
arbitrary dimension. Note that the fact that the maximal value is obtained 
on a qubit system is not trivial, although the measurements are dichotomic. 
For Kochen-Specker inequalities with dichotomic measurements examples are 
known, where the maximum value cannot be attained in a two-dimensional 
system \cite{dimwit} and also for Bell inequalities this has been observed
\cite{Pal10, moroder13}.

As a third application, we consider the noncontextuality scenario 
recently discovered by S. Yu and C.~H. Oh \cite{YuOh12}. There, thirteen 
measurements on a three-dimensional system are considered, and a 
noncontextuality inequality is constructed, which is violated by 
any quantum state. It has been shown that this scenario is the simplest 
situation where state-independent contextuality can be observed \cite{Cabello11}, so 
it is of fundamental importance. We can directly apply our method 
to the original inequality by Yu and Oh, as well as recent improvements 
\cite{KBLGC12} and compute the corresponding Tsirelson-like bounds. 
We recall that our results are not directly related to the phenomenon of quantum contextuality, since no compatibiliy of the measurements is assumed, but they show the effectiveness of our method even on complex scenarios, namely, inequalities containing 37 or 41 terms, that involve sequential measurements.
Our results are summarized in Table 1.
\begin{table}[t]
\begin{tabular}{|c|c|c|c|c|c|}
\hline
Ineq. & NCHV  & State-independent   & Algebraic & \textbf{Sequential}
\\
 & bound  & quantum value   & maximum & \textbf{bound}
\\\hline
Yu-Oh &  16 & $50/3\approx 16.67$   & 50 & 17.794\\
Opt2 & 16 & $52/3\approx 17.33$   & 52 &  20.287\\
Opt3 & 25 & $83/3\approx 27.67$   & 65 & 32.791
\\
\hline
\end{tabular}
\caption{\label{tab}%
Bounds on the quantum correlations for the Kochen-Specker inequalities 
in the most basic scenario. Three inequalities were investigated: First, 
the original inequality proposed in Ref.~\cite{YuOh12} and the optimal inequalities
from Ref.~\cite{KBLGC12} with measurement sequences of length two (Opt2) and 
 length three (Opt3). For each inequality, the following numbers are given:
the maximum value for noncontextual hidden variable (NCHV) models, the 
state-independent quantum violation in three-dimensional systems (obtained in Refs.~\cite{YuOh12,KBLGC12}), the algebraic maximum and the maximal 
value that can be attained in quantum mechanics for the sequential 
measurements. The latter bound  is higher than the
state independent quantum value, since the observables do not have to
obey the compatibility relations occurring in the Kochen-Specker theorem.
Notice that the sequential bound is obtained as a maximization over the set of possible observables and states, thus it is in general state-dependent.
Interestingly, in all cases the maximal quantum values are significantly below the
algebraic maximum.
}
\end{table}

Another  class of inequalities is given by the 
guess-your-neighbor's-input  inequalities \cite{GYNI10}, 
which if viewed as multipartite inequalities, show no quantum violation but
a violation with the use of postquantum no-signalling resources. 
We calculate the sequential bound for the case of measurement sequences of
length three, instead of measurement on three parties. We consider
\begin{equation}
\begin{split}
P(000|000) + P(110|011) + P(011|101) \\
+ P(101|110)\leq \Omega_{C,Q} \leq \Omega_S \leq \Omega_{NS},
\end{split}
\end{equation}
with the notation $P(r_1,r_2,r_3|s_1,s_2,s_3)$ as before, and possible results 
and settings $r_i\in \{0,1\}$ and $s_i\in\{0,1\}$. We find that 
\begin{equation}
\Omega_S \approx 1.0225,
\end{equation}
while it is known that $\Omega_{C,Q} = 1$ and $\Omega_{NS}=\tfrac{4}{3},$
where the indices $C,Q,S,NS$ label, respectively, the classical, quantum, 
sequential and no-signalling bounds. So, in this case, the bound for 
sequential measurements is higher than the bound for spatially separated 
measurements. This also highlights the greater generality of our method
in comparison with the results of Ref.~\cite{Fritz10}: there, only temporal
inequalities with sequences of length two have been considered, where in 
addition the measurements can be split in two separate groups. In this case 
it turned out that the bounds were always reached with commuting observables. 
Our examples show that this is usually not the case, when longer measurement 
sequences are considered.

{\it Discussion and conclusions.---}For interpreting our results, let us note that our scenario 
is  more general than the scenarios considered by 
Leggett and Garg and Kochen and Specker. Leggett and Garg consider 
a special time evolution $\vr(t) = U(t) \vr(0) U^\dagger(t),$ 
which is mapped in the Heisenberg picture onto the observables.
In our case, the observables are not connected via unitaries; 
this corresponds to a more general time evolution.
Compared with the Kochen-Specker scenario, our approach is more
general since it does not assume that the measurements in a 
sequence are commuting. Nevertheless,  if one wishes to connect 
existing noncontextuality inequalities to information processing 
tasks, it is important to know the maximal quantum values (also
if the observables do not commute), in order to characterize the 
largest quantum advantage possible.

Furthermore, we emphasize 
that in our 
derivation it was assumed that the measurements are described 
by projective measurements and this condition is indeed important. 
In fact, this sheds light on the role of projective measurements: 
one can easily construct classical devices with a memory, which 
give for sequential measurements as in Eq.~(\ref{klyachko}) the 
algebraic maximum $\mathcal{S}_5=5.$ These classical devices must 
also have a quantum mechanical description. Our results show, 
however, that in this quantum mechanical description more general 
than projective measurements must occur and a more general dynamical 
evolution than the  projection is required. From this perspective, 
our results prove that the memory that can be encrypted in quantum systems
by projective measurements is bounded.

Our results lead to the question of  why quantum mechanics does
not allow us to reach the algebraic maximum of temporal correlations, 
as long as projective measurements are considered. We believe that
proper generalizations of concepts such as information causality and 
communication complexity might play a role here, but we leave this
question for further research. A first step in explaining
quantum mechanics from information theoretical principles lies in the 
precise characterization of all possible temporal quantum correlations, 
and our work presents an operational solution to this problem.  

{\it Acknowledgements.---}We thank J.-D. Bancal, T. Fritz, Y.-C. Liang, G. Morchio, and M. Navascu\'es 
for discussions. This work has been supported by the 
EU (Marie Curie CIG 293993/ENFOQI) and the BMBF 
(Chist-Era Project QUASAR).

\appendix

\subsection{A1: Discussion of the simplest Leggett-Garg scenario}\label{A1}

In this part we provide some further details about how to determine the set of possible quantum values for the simplest non-trivial Leggett-Garg scenario as shown in Fig.~2 of the main text. Here it is assumed that one can measure an observable $M$ at three different time instances $t_1,t_2,t_3$ as shown in Fig.~1 of the main text, which gives rise to three different observables $A_i=M(t_i)$ with $i=1,2,3$. 

However, rather than being interested in determining the full sequential probability $P({\bf r | s})$ for all possible combinations we are here only interested in some limited information, namely only for the correlation space. This means that from a general distribution we only want to reproduce the correlations terms $\braket{A_iA_j}_{\rm seq}$ with  $1\leq i< j \leq 3$ each defined by 
\begin{equation}
\braket{A_iA_j}_{\rm seq} = P(r_i = r_j|i,j) - P(r_i \not= r_j|i,j).
\end{equation} 
Thus we want to characterize the set
\begin{align}
\nonumber
\mathcal{S}_{\rm qm} &= \{ q_{ij} \in \mathbbm{R}^3 : q_{ij} = \braket{A_i A_j}_{\rm seq},  \\ &  \braket{A_i A_j}_{\rm seq} \textrm{ has projective quantum rep.} \}.
\end{align}

For this we refer to problem given by Eq.~4 of the main text, with 
\begin{equation}
X \!= \!\left[ \begin{array}{ccc} 1 & \braket{A_1A_2}_{\rm seq} & \braket{A_1A_3}_{\rm seq} \\   \braket{A_1A_2}_{\rm seq} & 1 & \braket{A_2A_3}_{\rm seq} \\  \braket{A_1A_3}_{\rm seq} & \braket{A_2A_3}_{\rm seq} & 1 \end{array} \right].
\end{equation}
Any matrix of this form has a sequential projective quantum representation if 
and only if $X$ is positive semidefinite. However a matrix satisfies $X \succeq 
0$ if and only if the determinant of all principal minors are non-negative.  
This gives
\begin{align}
\nonumber
\mathcal{S}_{\rm qm} &= \{ q_{ij} \in \mathbbm{R}^3 : |q_{ij}| \leq 1 ,  \\ &  1 + 2 q_{12}q_{13}q_{23} \geq q_{12}^2 + q_{13}^2 + q_{23}^2 \}.
\end{align}
which is the plotted region of Fig.~2 of the main text.

We mention that via the general method one can also in principle determine the 
achievable probability distribution of a general scenario. However, this 
requires  the solution of a SDP with some unknown entries, and hence an 
analytic solution is in general not accessible.

\subsection{A2: Detailed discussion of bounds for the $N$-cycle inequalities}

We first need the general form \cite{AQBTC12} for Eq.~(10) of the main text
\begin{equation}
\mathcal{S}_N(\gamma) = \sum_{i=0}^{N-1}\gamma_i \mean{A_i A_{i+1}}_{\rm seq},
\label{ncycle2}
\end{equation}
where the indices are taken modulo $N$ and 
$\gamma=(\gamma_0,\ldots,\gamma_{N-1})\in \{-1,1\}^N$ with an odd number of 
$-1$.
Since any two assignments $\gamma$ and $\gamma'$ can be converted into each 
 other via some substitutions $A_i\rightarrow -A_i$, the quantum bound does not 
 depend on the particular choice of $\gamma$.
For the case odd $N$, we can consider the expression
\begin{equation}
\mathcal{S}_N = -\sum_{i=0}^{N-1} \langle A_i A_{i+1}\rangle_\mathrm{seq},
\end{equation}
with index $i$ taken modulo $N$. 
The optimization problem in Eq. (4) of the main text, therefore, can be expressed as
\begin{equation}\label{ncsdp}
\begin{split}
\text{maximize:}&\quad \frac{1}{2}\tr(WX)\\
\text{subjected to:}&\quad X=X^T\succeq 0\ \text{and}\ X_{ii}=1\ \text{for all 
} i,
\end{split}
\end{equation}
where $W$ is the circulant symmetric matrix
\begin{equation}\label{W}
W=- \left[
\begin{array}{ccccc} 0    & 1 & \dots  & 0 & 1  \\
1 & 0    & 1 &       \  & 0  \\\vdots  & 1& 0    & \ddots  & \vdots   \\
0  &        & \ddots & \ddots  & 1   \\1  & 0 & \dots  & 1 & 0 \\\end{array}\ 
\right] \ .
\end{equation}
The condition $X\succeq 0$, i.e. $v^T X v\geq$ for any real vector $v$, 
follows from the fact that 
$\mean{A_iA_j}_{\rm seq}  = \frac{1}{2}\tr[\varrho (A_i A_j + A_j A_i)]$
and the fact that the matrix $Y=\tr[\varrho (A_i A_j)]$ fulfils 
$v^T Y v\geq$ for any real vector $v$, and $X$ is the real part of 
$Y$.

By using the vector ${\lambda=(\lambda_1,\ldots, \lambda_N)}$, the dual 
problem for the semidefinite program in Eq.~(\ref{ncsdp}) can be written 
as (see Ref.~\cite{sdpref} for a general treatment and Ref.~\cite{WEH06} 
for the discussion of a similar problem)
\begin{equation}\label{ncsdpD}
\begin{split}
\text{minimize:}&\quad \tr(diag(\lambda))\\
\text{subjected to:}&\quad -\frac{1}{2}W+diag(\lambda)\succeq 0,
\end{split}
\end{equation}
where $diag(\lambda)$ denotes the diagonal matrix with entries $\lambda_1,\ldots,\lambda_N$.

Let us denote with $p$ and $d$ optimal values for, respectively, the primal 
problem in Eq.~\eqref{ncsdp} and the dual problem in Eq.~\eqref{ncsdpD}. Then 
$d \geq p$.  We shall provide a feasible solution for the dual problem with  
$d=N \cos(\frac{\pi}{N})$ and a feasible solution for the primal problem with 
$p=d$, this will guarantee the optimality of our primal solution.

We start by finding the maximum eigenvalue for $W$. Since $W$ is a circulant matrix, its eigenvalues 
can be written as \cite{Gray}
\begin{eqnarray}
 \mu_j = -2 \cos\left(\frac{2\pi j}{N}\right)
\end{eqnarray}
for $j=0,\ldots, N-1$, and ${ \mu_{max}= 2\cos\left(\frac{\pi}{N}\right) }$ the maximum eigenvalue.

For a pair of Hermitian matrices $A,B$, it holds $\mu_{min}(A+B)\geq \mu_{min}(A)+\mu_{min}(B)$, where $\mu_{min}$ denotes the minimum eigenvalue. Therefore, $\lambda=(\cos\left(\frac{\pi}{N}\right), \ldots, \cos\left(\frac{\pi}{N}\right) )$ is a 
feasible solution for the dual problem and $\tr[diag(\lambda)]= N \cos\left(\frac{\pi}{N}\right)$, 
and $p \leq N \cos\left(\frac{\pi}{N}\right)$. 

Now consider the matrix $X'_{ij}= (x_i, x_j)$, with $x_1,\ldots,x_N$ unit vectors in a 
2-dimensional space such that the angle between $x_i$ and $x_{i+1}$ is 
$\frac{N+1}{N}\pi$, and $(\cdot,\cdot)$ denoting the scalar product. Clearly, 
$X'$ is positive semidefinite.
Since $X'_{i,i+1}= -\cos\left(\frac{\pi}{N}\right)$, it follows that $p=d=N 
\cos\left(\frac{\pi}{N}\right)$ and the solution $X'$ is optimal.

In order to prove that $X'$ can be obtained as matrix of expectation values for 
sequential measurements, we define for a 3-dimensional unit vector $\vec{a}$ 
the observable $\sigma_a\equiv \vec\sigma\cdot \vec a$, where $\vec \sigma$ 
denots the vector of the Pauli matrices.
Then, by Eq.~(3) of the main text, $\langle \sigma_a 
\sigma_b\rangle_{seq}=\vec{a}\cdot\vec{b}$, independently of the initial 
quantum state $\varrho$. In fact, explicit observables
reaching this bound have already been discussed in the literature \cite{Barbieri09,dimwit}.

For the case $N$ even, we can consider the expression
\begin{equation}
\mathcal{S}_N = \sum_{i=0}^{N-2} \langle A_i A_{i+1}\rangle_{seq} - \langle A_0 A_{N-1}\rangle_{seq},
\end{equation}
and the maximization problem can be expressed as a SDP as in Eq.~(\ref{ncsdp}), with the proper choice of the matrix $W$. Such a SDP has been solved in Ref.~\cite{WEH06}. The solution is analogous to the previous one: A set of observables, for a two-level system, saturating the bound, again, independently of the quantum state, is given by observables $A_i=\vec{\sigma} \cdot \vec{x_i}$, where the vectors $x_i$ are on a plane with an angle $\frac{\pi}{N}$ separating $x_i$ and $x_{i+1}$.

As opposed to the $N$ odd case, such a bound can be also reached with commuting operators, this corresponds to the well known maximal violation of Braunstein-Caves inequalities~\cite{WEH06}.

The above results prove that the bound computed in Ref.~\cite{dimwit} for sequential measurements on qubits, coinciding with the value explicitly obtained in Ref.~\cite{Barbieri09}, is valid for any dimension of the quantum system on which measurements are performed.  

Finally, we stress that the construction of the above set of observables from 
the  solution of the SDP, i.e., the matrix $X$ or the set of vectors $\{x_i\}$ 
such that $X_{ij}=(x_i,x_j)$, is general.  We recall that the vectors $\{x_i\}$ 
can be obtained, e.g., as the columns of the matrix $\sqrt{X}$ and, therefore, 
the dimension of the subspace spanned by them is equal to the rank of the 
matrix $X$.
In the previous case, since we were dealing with vectors in dimension $d\leq 3$, we used the property of Pauli matrices
\begin{equation}\label{cliff}
\{ \sigma_a ,\sigma_b\}\equiv \sigma_a \sigma_b + \sigma_b \sigma_a=2(\vec{a}\cdot\vec{b})\id.
\end{equation}

For matrices $X$ with higher rank, the corresponding vectors $\{x_i\}$ will 
span a real vector space $V$ of dimension $d > 3$.  Now for general complex 
vector spaces $V$ with a symmetric bilinear form $(~,~)$, an analogue of 
Eq.(\ref{cliff}), namely
\begin{equation}\label{cliffn}
\{ A_v ,A_u\}=2(v,u)\id, \ \text{ for any } u,v\in V
\end{equation}
 can be established by a representation of associated Clifford algebra, cf.\ 
Ref.~\cite{Tsi85,Tsi87}

As a consequence, for every positive semidefinite real matrix $X$ with diagonal elements equal to $1$, one can find a set of unit vectors $\{x_i\}$ giving $X_{ij}=(x_i,x_j)$ and a set of $\pm1$-valued observables $\{A_i\}$, associated with $\{x_i\}$ , such that
\begin{equation}
\mean{A_i A_j}_{seq} = \tr\left[\frac{1}{2}\varrho (A_i A_j + A_j A_i)\right] =(x_i,x_j),
\end{equation}
for all quantum states $\varrho$. In particular, if the rank of $X$ is $d$, such operators can be chosen as $2^d\times 2^d$ Hermitian matrices \cite{BraRob}. 
This shows the completeness of the simplified method.

\subsection{A3: Completeness of the general method}\label{A3}

In this part we shortly comment on the completeness of the presented general method. As pointed out, this has already been proven indirectly in the context of the spatial bipartite case~\cite{NPA08}.

At first let us change slightly the notation in order to make it closer to the 
one used in Ref.~\cite{NPA08}. In the following we do not explicitly consider 
the matrix $M$ from the main text, but rather a slightly smaller matrix where 
one erases some trivial constraints. In the following the set $\{ E_i \} $ 
contains all projectors $\Pi_k^s$, but one of the outcomes $k$ from each 
setting $s$ is left out. We also use a single subscript to identify setting and 
outcome. Then the matrix
\begin{equation}
\chi^n_{{\bf uv}} = \tr[E({\bf u})E({\bf v})^\dag \rho]
\end{equation}
with ${\bf u} = (u_1,u_2,\ldots,u_l)$ is built from all products $E({\bf u}) = 
E_{u_1} E_{u_2} \cdots E_{u_l}$ of the operators $\{E_i\}$ of at most length 
$l\leq n$, and the single extra ``sequence'' ${\bf u}= 0$ of the identity 
operator, $E(0) = \mathbbm{1}$. Again this matrix has to satisfy linear 
relations parsed as $\chi^n_{\bf uv} = \chi^n_{\bf u^\prime v^\prime}$, if the 
operators fulfill $E({\bf u})E({\bf v})^\dag =  E({\bf u^\prime})E({\bf 
v^\prime})^\dag$ as a consequence of the orthogonality properties of 
projectors, and that $\chi^n \succeq 0$.

That this matrix is positive semidefinite can be verified as follows: Let us first assume that there exists a sequential projective quantum representation. Consider the operator $C = \sum_{\bf u} c_{\bf u} E({\bf u})^\dag$ with arbitrary $c_{\bf u} \in \mathbbm{C}$ and evaluate the expectation value of $CC^\dag$, which provides
\begin{align}
\tr(CC^\dag \varrho) &= \sum_{{\bf u,v} } c_{\bf u} \tr[E(\mathbf u)^\dag 
E(\mathbf v) \varrho] c_{\bf v}^* \\
\label{eq:chi_positivity}
& = \sum_{{\bf u,v} } c_{\bf u} \chi^n_{{\bf uv}} c_{\bf v}^* \geq 0.
\end{align}
The final inequality holds because $CC^\dag\succeq 0$ and $\rho\succeq 0$ are 
both positive semidefinite operators. Since $c_{\bf u} \in \mathbbm{C}$ are 
arbitrary the condition given by Eq.~(\ref{eq:chi_positivity}) means that 
$\chi^n\succeq 0$ is positive semidefinite.

For the reverse one needs a way to construct an explicit sequential projective 
quantum representation out of the matrix $\chi^n$ satisfying the above 
properties. For this, clearly more difficult part, we refer to 
Ref.~\cite{NPA08} and just mention the solution. For the given positive 
semidefinite matrix $\chi^n$ one associates a set of vectors $\{ \ket{e_{\bf 
u}} \}$ by the relation $\chi^n_{\bf uv}=\braket{e_{\bf u}|e_{\bf v}}$. From 
this set of vectors one now constructs an appropriate state and corresponding 
projective measurements by $ \hat{\mathcal{H}} = \Span(\{ \ket{e_{\bf u}} \})$, 
$\hat \rho = \ket{e_0}\bra{e_0}$, and $\hat E_i = \Proj( \Span(\{ \ket{e_{\bf 
u}}: u_1 = i\}))$ where $\Proj$ means the projector onto the given subspace.  
That these solution satisfies all the required constraints is shown in the 
proof of Theorem 8 of Ref.~\cite{NPA08}. An analogous mathematical result, valid only for the
case of dichotomic observables, has been presented also in Ref.~\cite{PNA10}.

In the spatial case considered in Ref.~\cite{NPA08},  some of these operators, 
additionally, have to commute  since they should correspond to measurements 
onto different local parts. This cannot be inferred, in general, by a finite 
level  $\chi^n$ and this is eventually the reason why in the spacial case 
arbitrary high order terms have to be considered. However, luckily, since in 
our situation the measurements of different settings may well fail to commute 
we can rely on a finite level $n$.

\end{document}